\documentstyle[12pt]{article}

\title { B\"acklund transformation for the Krichever-Novikov equation }
\author{                         V.E. Adler                           }
\date  {                        18 July 1997                          }

\def\be{\begin{equation}}
\def\ee{\end{equation}}
\def\qed{\vrule height0.7em width0.4em depth0pt}
\def\const{\mbox{const}}
\begin{document}
\maketitle

The Krichever-Novikov equation
\be u_t= u_{xxx}-{3\over2u_x}(u^2_{xx}-r(u))+cu_x, \quad r^{(5)}=0
\label{ut}\ee
appeared (up to change $u=p(\tilde u),\ \dot p^2=r(p)$) in \cite{kn} for the
first time in connection with study of finite-gap solutions of the
Kadomtsev-Petviashvili equation.  The distinctive feature of the equation
(\ref{ut}) is that, accordingly to \cite{ssy}, no differential substitution
exists connecting it with other KdV-type equations.  This property impedes
the construction of the B\"acklund transformation (BT) which in other cases
can be obtained by composition of two differential substitutions.
Nevertheless, we demonstrate that (\ref{ut}) admits BT which connects it with
other equation of the same form
\be v_t= v_{xxx}-{3\over2v_x}(v^2_{xx}-R(v))+cv_x, \quad R^{(5)}=0.
\label{vt}\ee

\paragraph{Theorem 1.}  Let $h(u,v)$ be arbitrary polynomial of degree less
than 3 on each variable, and $r(u)$ and $R(v)$ be its discriminants as
quadratic trinomial on $v$ and $u$ correspondingly:
\be h_{uuu}=h_{vvv}=0, \quad r(u)= h^2_v-2hh_{vv},\quad R(v)= h^2_u-2hh_{uu}.
\label{hrr}\ee
Then formula
\be u_xv_x = h(u,v) \label{uvx}\ee
defines BT between equations (\ref{ut}) and (\ref{vt}). \qed
\bigskip

It is easy to see that relation $h(u,v)=0$ gives a birational transformation
of the curves
\be y^2=r(u), \quad Y^2=R(v). \label{yY}\ee
The form of the relations (\ref{ut}) --- (\ref{uvx}) does not change under
arbitrary nondegenerate linear-fractional substitutions
$$ u={k\bar u+\ell\over m\bar u+n}, \quad
   v={K\bar v+   L\over M\bar v+N}. $$

It is clear that polynomials $r$ and $R$ cannot be chosen independently
(indeed, they contain 10 coefficients, while $h$ only 9).  From the other
hand, if equations (\ref{hrr}) are solvable with respect to $h,$ then their
solution can be not unique and then question arises about permutability of
two different BT.

We consider only the case $r(u)=R(u)$ and assume that zeroes of $r$ are
simple.  Accordingly to \cite{ash} it is possible if and only if the
polynomial $h(u,v)$ is symmetric and irreducible.  It should be noted, that
reducible cases are also rather meaningful, for example the formula
(\ref{uvx}) where $r=0,\ h=(v-\mu u)^2$ defines the BT for Schwartz-KdV
equation, from which the BT for other KdV-type equations can be derived by
standard substitutions.

The equation $h(u,v)=0$ is known as Euler-Chasles correspondence (see e.g.
\cite{bv}) and is equivalent to the shift on the elliptic curve, which we
take, without loss of generality, in the Weierstrass form:
$r(u)=4u^3-g_2u-g_3.$  The corresponding polynomial $h$ depends on arbitrary
parameter $\mu$ and is of the form $h=H(u,v,\mu)/\sqrt{r(\mu)},$ where
\be H(u,v,\mu) = (uv+\mu u+\mu v+{g_2\over4})^2 - (u+v+\mu)(4\mu uv-g_3).
\label{H}\ee
The relation $H=0$ where $u=\wp(z),\ v=\wp(z\pm a),\ \mu=\wp(a)$ is nothing
but Euler form of the addition theorem for Weierstrass $\wp$-function.  Hence
it is evident, that composition of the correspondences $H(u,v,\mu)=0,\
H(v,w,\nu)=0$ coincides with composition $H(u,\tilde v,\nu)=0,\ H(\tilde
v,w,\mu)=0.$  Indeed, in the both cases $w$ takes 4 possible values $\wp(z\pm
a\pm b).$

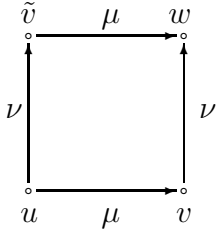
\begin{figure}[h]
\setlength{\unitlength}{0.05em}
 \begin{picture}(150,140)(-50,-25)
   \put(0,100){\circle{4}}                \put(100,100){\circle{4}}
   \put(-5,108){$\tilde v$}    \put(90,108){$w$}
                 \put(5,100){\vector(1,0){90}}
                 \put(47,108){$\mu$}
   \put(0,5){\vector(0,1){90}}            \put(100,5){\vector(0,1){90}}
   \put(-15,47){$\nu$}                    \put(110,47){$\nu$}
                 \put(5,0){\vector(1,0){90}}
                 \put(47,-20){$\mu$}
   \put(0,0){\circle{4}}                  \put(100,0){\circle{4}}
   \put(-5,-20){$u$}                       \put(95,-20){$v$}
 \end{picture}
\end{figure}

The Euler-Chasles correspondence is formally a particular case of (\ref{uvx})
for $u=\const$ (but it should be noted, that the constant solutions of the
equation (\ref{ut}) are exhausted by zeroes of $r$).  It turns out that
analogous commutativity property is valid for the BT (\ref{uvx}) itself.  Let
$u,v,\tilde v,w$ are related by the BT as on the diagram above, that is
\be \begin{array}{ccc}
 u_xv_x=H(u,v,\mu)/\alpha, &
 v_xw_x=H(v,w,\nu)/\beta,  &
 \alpha^2=r(\mu), \\
 u_x\tilde v_x=H(u,\tilde v,\nu)/\beta,  &
 \tilde v_xw_x=H(\tilde v,w,\mu)/\alpha, &
 \beta^2=r(\nu).
\end{array} \label{dx}\ee
Let us remind that the constraint $P(u,v,\tilde v,w)=0$ is called the
nonlinear superposition principle if it is nondegenerate on each variable and
its derivative on $x$ vanishes in virtue of itself: $D_x(P)|_{P=0}=0.$

\paragraph{Theorem 2.}  Nonlinear superposition principle for the BT
(\ref{uvx}), (\ref{H}) is given by formula
$$ P = k_0uvw\tilde v -k_1(uvw+vw\tilde v+w\tilde vu+\tilde vuv)
      +k_2(uw+v\tilde v) - $$
$$ \qquad -k_3(uv+\tilde vw)-k_4(u\tilde v+vw)+k_5(u+v+w+\tilde v)+k_6 = 0 $$
where
$$ k_0= \alpha     +\beta,             \quad
   k_1= \alpha\nu  +\beta\mu,          \quad
   k_2= \alpha\nu^2+\beta\mu^2,        \quad
   k_5= {g_3\over2}k_0+{g_2\over4}k_1, \quad
   k_6= {g^2_2\over16}k_0+g_3k_1,      $$
$$ k_3= {\alpha\beta(\alpha+\beta)\over2(\nu-\mu)}
        - \alpha\nu^2 + \beta(2\mu^2-{g_2\over4}),  \qquad
   k_4= {\alpha\beta(\alpha+\beta)\over2(\mu-\nu)}
        - \beta\mu^2 + \alpha(2\nu^2-{g_2\over4}).  \qquad\qed $$
\smallskip

Note that elimination of derivatives from (\ref{dx}) yields the relation
$$ S = \alpha^2H(u,\tilde v,\nu)H(v,w,\nu) -
       \beta^2H(u,v,\mu)H(\tilde v,w,\mu) = 0, $$
which is not the nonlinear superposition principle by itself.  It is
explained by reducibility of the polynomial $S=PQ,$ where $P$ is written
above and $Q$ is obtained from $P$ by substitution $\beta$ to $-\beta.$  In
contrast to the constraint $P=0$ the constraint $Q=0$ is not compatible with
dynamics on $x.$

\medskip

The work was supported by grants INTAS-93-166-Ext and RFBR-96-01-00128.


\end{document}